\documentclass[aps,pra,superscriptaddress,twocolumn,amssymb,floatfix]{revtex4-1}

\usepackage{color}
\usepackage{amsmath}
\usepackage{amsthm}
\usepackage{amssymb}
\usepackage{amsfonts}
\usepackage{amstext}
\usepackage{graphicx}	
\usepackage{bbm}
\usepackage{enumerate}
\usepackage{tikz}
\usepackage{xr}	
\usepackage{cancel}

\usepackage[colorinlistoftodos,prependcaption,textsize=tiny]{todonotes}

\newcommand{\ket}[1]{ \left\lvert #1\right\rangle}

\newcommand{\Pur}[1]{\mathcal{P}\left( #1 \right)}
\newcommand{\tr}[1]{\mathrm{tr}\left( #1 \right)}

\newcommand{\setA}{\ensuremath{\mathcal{A}}}

\newcommand{\setC}{\ensuremath{\mathcal{C}}}
\newcommand{\setS}{\ensuremath{\mathcal{S}}}

\begin{document}

\title{Multipartite Entanglement Analysis From Random Correlations}

\author{Lukas Knips}
\email{lukas.knips@mpq.mpg.de}
\affiliation{Max-Planck-Institut f\"{u}r Quantenoptik, Hans-Kopfermann-Stra{\ss}e 1, 85748 Garching, Germany}
\affiliation{Department f\"{u}r Physik, Ludwig-Maximilians-Universit\"{a}t, Schellingstra{\ss}e 4, 80799 M\"{u}nchen, Germany}
\affiliation{Munich Center for Quantum Science and Technology (MCQST), Schellingstra{\ss}e 4, 80799 M\"{u}nchen, Germany}

\author{Jan Dziewior}
\affiliation{Max-Planck-Institut f\"{u}r Quantenoptik, Hans-Kopfermann-Stra{\ss}e 1, 85748 Garching, Germany}
\affiliation{Department f\"{u}r Physik, Ludwig-Maximilians-Universit\"{a}t, Schellingstra{\ss}e 4, 80799 M\"{u}nchen, Germany}
\affiliation{Munich Center for Quantum Science and Technology (MCQST), Schellingstra{\ss}e 4, 80799 M\"{u}nchen, Germany}

\author{Waldemar K{\l}obus}
\affiliation{Institute of Theoretical Physics and Astrophysics, Faculty of Mathematics, Physics and Informatics, University of Gda\'nsk, 80-308 Gda\'nsk, Poland}

\author{Wies{\l}aw Laskowski}
\email{wieslaw.laskowski@ug.edu.pl}
\affiliation{Institute of Theoretical Physics and Astrophysics, Faculty of Mathematics, Physics and Informatics, University of Gda\'nsk, 80-308 Gda\'nsk, Poland}
\affiliation{International Centre for Theory of Quantum Technologies, University of Gda\'nsk, 80-308 Gda\'nsk, Poland}

\author{Tomasz~Paterek}
\affiliation{Institute of Theoretical Physics and Astrophysics, Faculty of Mathematics, Physics and Informatics, University of Gda\'nsk, 80-308 Gda\'nsk, Poland}
\affiliation{School of Physical and Mathematical Sciences, Nanyang Technological University, 637371 Singapore}
\affiliation{MajuLab, International Joint Research Unit UMI 3654, CNRS, Universit\'{e} Cote d'Azur, 
Sorbonne Universite, National University of Singapore, Nanyang Technological University, Singapore. }

\author{Peter J. Shadbolt}
\affiliation{Department of Physics, Imperial College London, London SW7 2AZ, United Kingdom}

\author{Harald Weinfurter}
\affiliation{Max-Planck-Institut f\"{u}r Quantenoptik, Hans-Kopfermann-Stra{\ss}e 1, 85748 Garching, Germany}
\affiliation{Department f\"{u}r Physik, Ludwig-Maximilians-Universit\"{a}t, Schellingstra{\ss}e 4, 80799 M\"{u}nchen, Germany}
\affiliation{Munich Center for Quantum Science and Technology (MCQST), Schellingstra{\ss}e 4, 80799 M\"{u}nchen, Germany}

\author{Jasmin D. A. Meinecke}
\affiliation{Max-Planck-Institut f\"{u}r Quantenoptik, Hans-Kopfermann-Stra{\ss}e 1, 85748 Garching, Germany}
\affiliation{Department f\"{u}r Physik, Ludwig-Maximilians-Universit\"{a}t, Schellingstra{\ss}e 4, 80799 M\"{u}nchen, Germany}
\affiliation{Munich Center for Quantum Science and Technology (MCQST), Schellingstra{\ss}e 4, 80799 M\"{u}nchen, Germany}

\begin{abstract}
	Quantum entanglement is usually revealed via a well equationed, carefully chosen set of measurements.
	Yet, under a number of experimental conditions, for example in communication within multiparty quantum networks, noise along the channels or fluctuating orientations of reference frames may ruin the quality of the distributed states.
	Here we show that even for strong fluctuations one can still gain detailed information about the state and its entanglement using random measurements.
	Correlations between all or subsets of the measurement outcomes and especially their distributions provide information about the entanglement structure of a state. 
	We analytically derive an entanglement criterion for two-qubit states and provide strong numerical evidence for witnessing genuine multipartite entanglement of three and four qubits.
	Our methods take the purity of the states into account and are based on only the second moments of measured correlations.
	Extended features of this theory are demonstrated experimentally with four photonic qubits.
	As long as the rate of entanglement generation is sufficiently high compared to the speed of the fluctuations, this method overcomes any type and strength of localized unitary noise.
\end{abstract}

\maketitle

\section*{Introduction}

One of the most striking features of quantum entanglement is the existence of correlated measurement outcomes between spatially separated particles, which exceed expectations based on classical physics.
These correlations are typically observed with carefully equationed local measurements.
They get distorted if a common reference frame is lacking and especially in the presence of noise along the channels distributing the entangled particles.
In practice, for many channels the instabilities are often irremovable: optical fibers rotate polarization, changing phases affect a path degree of freedom, atmospheric turbulence acts on the modes of orbital angular momentum, magnetic field fluctuations influence trapped ions, etc.
Common sense tells that this renders the distributed quantum state useless and unrecognizable.

Here we provide a method for entanglement detection and analysis that is insensitive to local rotations and thus overcomes these difficulties.
It requires neither reference frames nor equationment nor calibration of measuring devices.
Still, it can both witness as well as classify multipartite entanglement in the presence of local unitary noise.
The key to overcome the lack of control and knowledge regarding each single measurement is to harness uniform sampling of the entirety of all measurements.
Especially without any prior knowledge about the state, the conceptually simple method of random sampling proves highly beneficial for entanglement detection and state analysis.

Previous work on entanglement detection relaxing the requirement of fully equationed reference frames first considered the absence of a shared reference frame, but still required the ability to choose or at least to repeat local measurement settings from a given set in order to detect, for example, the violation of a Bell inequality~\cite{PhysRevA.83.022110,PhysRevA.85.024101,PhysRevA.86.032322,PhysRevA.91.052118,PhysRevLett.104.050401,SciRep.2.470}, or for tomographic reconstruction~\cite{PhysRevA.99.052323}.
Under the same constraints, also adaptive methods for entanglement detection have been developed~\cite{PhysRevLett.108.240501,PhysRevA.88.022327}.
In the absence of any reference frames Bell violations can be measured with some probability~\cite{PhysRevA.96.012101,PhysRevA.98.042105} and entanglement can be detected by evaluating the second moment of the distribution of correlations obtained by measuring random observables on each subsystem~\cite{QuantInfComp.8.773,PhysRevA.77.062334,PhysRevA.80.042302,PhysRevA.80.042302,PhysRevA.90.042336,PhysRevA.92.050301,PhysRevA.94.042302}. Furthermore, it has been shown recently that higher-order moments of this distribution allow discrimination of very specific types of multipartite entanglement~\cite{arXiv.1808.06558}.
While these methods analyze full correlations, a recent experiment used second moments of subsets to deduce entanglement in systems of more than ten particles~\cite{Brydges260}.
In contrast, here we are interested in the detection of genuine multipartite entanglement, i.e. revealing that all particles share quantum entanglement.

We qualitatively investigate not only a specific moment of the distributions of full correlations, but all probability distributions of full as well as of marginal correlations, taking into account their interdependencies.
We show how they provide a detailed picture of the type of state and its entanglement structure for certain examples of pure states.
This illuminates the way to derive a general witnesses of genuine multipartite entanglement for arbitrary pure and mixed states.
These witnesses retain simplicity, as they are based only on second order moments of the distributions, and yet they outperform other criteria based on second moments~\cite{QuantInfComp.8.773,PhysRevA.77.062334,PhysRevA.80.042302,PhysRevA.80.042302,PhysRevA.90.042336,PhysRevA.92.050301,PhysRevA.94.042302}. 
We experimentally measure full and marginal distributions of correlations for various multiqubit states using reference frame free random measurements and show the applicability of all our extended analysis methods.
These methods are robust as they do not depend on the local unitary noise as long as the rate of generated entangled states is high enough to estimate the correlations for a momentarily constant noise.

\section*{Results}
\subsection*{Scenario}
Consider a source producing copies of an unknown $n$-qubit state $\varrho$, which is transmitted through unstable quantum channels to $n$ local observers (Fig.~1).
\begin{figure}
	\includegraphics[width=0.48\textwidth]{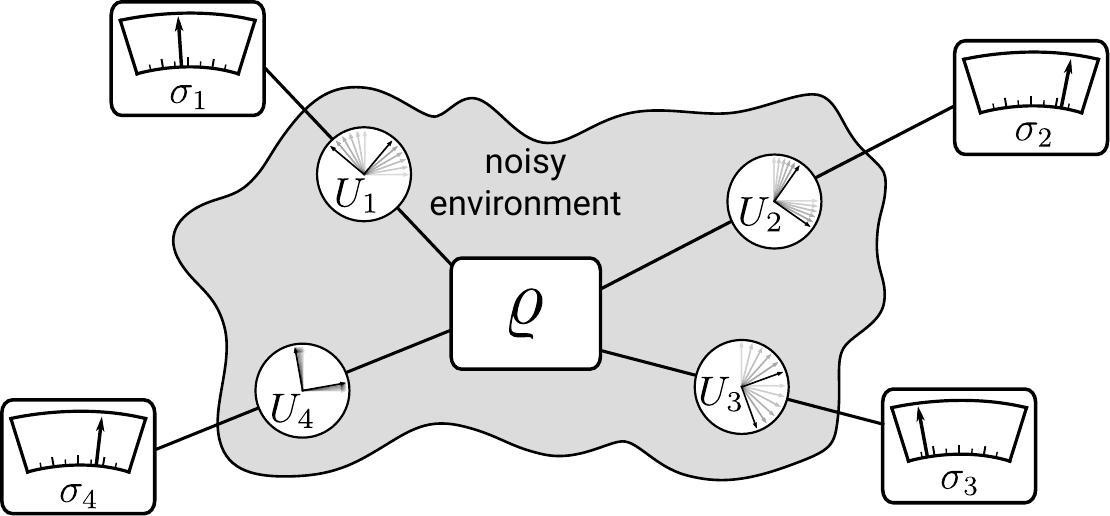}
	\caption{Quantum communication over noisy channels.
		A source produces an entangled state of, say, four qubits. 
		Each of them propagates through a noisy channel resulting in an unknown unitary transformation.
		When choosing local observables $\sigma_i$ uniformly at random, the statistics of correlations reveal detailed information on multipartite entanglement, independently of the noise in the channels or of the lack of shared reference frames.}
\end{figure}
During the $j$-th transmission the state $\varrho$ is transformed by $n$ random local unitary operators $U^{(j)}_i$ with $i=1,2,\dots,n$ according to
\begin{equation}
\varrho \rightarrow \varrho^{(j)} = U^{(j)}_1\otimes \ldots \otimes U^{(j)}_n \; \varrho \; U^{(j)\dagger}_1 \otimes \ldots \otimes U^{(j) \dagger}_n.
\label{eqn:transformedRho}
\end{equation}
Additionally, each of the $n$ observers is free to choose an arbitrary measurement setting $\sigma^{(j)}_i$ to measure her qubit.
If for each transmitted copy of $\varrho$ the transformations $U^{(j)}_i$ change significantly, all information about the state is lost.
However, in a very common scenario encountered by experimenters the unitary noise has a timescale which is sufficiently slow to obtain at least a few copies of $\varrho$ which have been affected by essentially the same noise, i.e., by 
the same set of local transformations $U^{(j)}_i$.
In this case the transformations are still much too fast to apply standard techniques of state analysis~\cite{PhysRevA.64.052312}, yet, it becomes possible to use the few equally transformed states to reliably record correlations
\begin{eqnarray}
E^{(j)}_{1\dots n} & = & \tr{\sigma^{(j)}_1\otimes \sigma^{(j)}_2  \otimes \ldots \otimes \sigma^{(j)}_n \; \varrho^{(j)}} \nonumber \\
& = & \tr{\tilde{\sigma}^{(j)}_1 \otimes \tilde{\sigma}^{(j)}_2  \otimes \ldots \otimes \tilde{\sigma}^{(j)}_n \; \varrho},
\label{eqn:fullcorrelations}
\end{eqnarray}
where each observer is keeping her local observable $\sigma^{(j)}_i$ constant in the timescale of constant noise, which results in the effective random observable $\tilde{\sigma}^{(j)}_i \equiv U^{(j) \dagger}_i \; \sigma_i \; U^{(j)}_i$.
Note that here and below the index $j$ refers to a set of transmitted states which have all been affected by the same noise transformations and measured using the same settings.

We refer to $E^{(j)}_{1\dots n}$ as ``full correlation'' or $n$-partite correlation because it involves measurement outcomes of all $n$ observers.
Besides full correlations, also ``marginal correlations'' can be measured, which are computed from the outcomes of a subset of observers.
For example, the marginal correlation of all observers but the first one is
\begin{equation}
E^{(j)}_{2\dots n} = \tr{\mathbbm{1} \otimes \tilde{\sigma}^{(j)}_2 \otimes \dots \otimes \tilde{\sigma}^{(j)}_n \; \varrho}.
\label{eqn:marginalcorrelations}
\end{equation}

The essential ingredient in our approach is that each observer samples local measurement directions $\tilde{\sigma}^{(j)}_i$ randomly according to a Haar uniform distribution. 
This removes any dependence of the obtained information on the actual structure or time dependence of the various $U^{(j)}_i$ and thus overcomes any bias in the random noise.

In our experiment we prepare four different four-qubit states using entangled photon pairs, where we encode two qubits in the polarization degree of freedom and two qubits in the path degree of freedom.
To comprehensively demonstrate the informational content of distributions of random correlations, we consider four quantum states belonging to different entanglement classes, in particular a tri-separable, a bi-separable and two genuinely multipartite entangled states, namely a Greenberger-Horne-Zeilinger (GHZ) state and a cluster state,

\begin{subequations}
\begin{eqnarray}
\ket{\psi_{\mathrm{trisep}}}&\propto&\left(\ket{00}+\ket{11}\right)\otimes\ket{0}\otimes\ket{0}, \label{eq:trisep} \\
\ket{\psi_\mathrm{bisep}}&\propto&\left(\ket{00}+\ket{11}\right)\otimes\left(\sin{\varphi}\ket{00}+\cos{\varphi}\ket{11}\right), \label{eq:bisep}\\
\ket{\rm GHZ}&\propto&\left(\ket{0000}+\ket{1111}\right), \label{eq:GHZ}\\
\ket{\setC_4}&\propto&\left(\ket{0000}+\ket{0011}-\ket{1100}+\ket{1111}\right).\label{eq:cluster}
\end{eqnarray}\label{EQ_STATES}
\end{subequations}
We utilize the full experimental control over the choice of measurement settings to emulate the local unitary transformations due to noisy channels and the Haar random choices of measurement settings.

Our experimental setup is based on spontaneous parametric down conversion, generating a pair of polarization entangled photons.
Those photons are sent to two Sagnac interferometers with polarizing beam splitters, adding a path degree of freedom, which is then coupled inside the interferometer with the polarization of the incoming photon.
This way, the two photons effectively provide four qubits.
Local transformations of the polarization inside the interferometer, which translate to path transformations behind the second polarizing beam splitter, together with polarization transformations outside the interferometer allow to locally modify and analyze both path and polarization degrees of freedom of both photons.
Further details of the setup can be found in \cite{Knips2016}.
It should be noted that while we clearly can deduce how characteristics of the state are reflected in the form of the measured distributions the other direction of deduction is in general much more difficult.

\subsection*{Analyzing Entanglement Structures}

\begin{figure*}
	\includegraphics[width=0.98\textwidth]{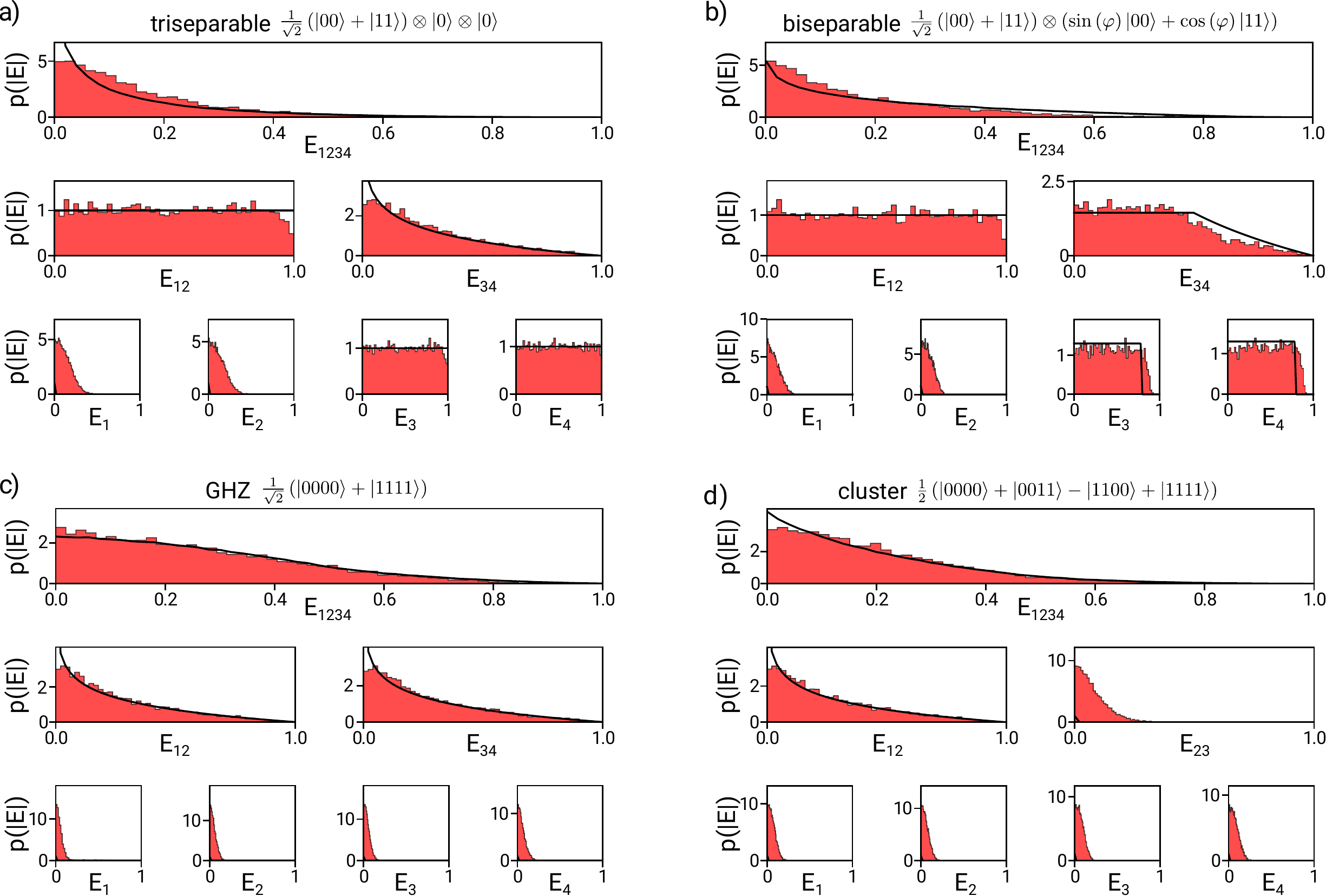}
	\caption{Experimental distributions of correlations for four typical states.
		For each state we plot the distribution $p(E)$ of the modulus of the measured full correlation $E_{1234}$ together with two of the six two-qubit marginal distributions and all four single-qubit marginals. 
		For this visualization, we measured each state along $10^4$ different settings (in panel b only $6000$ settings; we choose $\varphi \approx 0.2$).
		The histograms ($50$ bins) are derived from raw measured data corrected for detection efficiencies. 
		Solid lines represent theoretical curves for ideal states. 
		Deviations of the measured data from the ideal distributions are due to finite statistics and finite fidelity of the state preparation.}
\end{figure*}

In the following we study distributions of random correlations for these four states, see Fig.~2.
It is helpful to recall that for some particular states the distributions are known analytically.
A pure product state of $n$ qubits results in a distribution proportional to $-(\mathrm{ln}|E|)^{n-1}$ \cite{PhysRevA.92.050301,PhysRevA.94.042302}, which becomes uniform for $n=1$,
and a maximally entangled state of two qubits gives rise to a flat distribution~\cite{phdthesis_jdam,ShadboltSpringer}.
In addition to this established knowledge, we use new criteria to show that the experimental data not only provide information about the amount of entanglement in the full state, 
but also give insight into how the entanglement is shared among the parties, allowing to reconstruct the whole multipartite entanglement structure.
An important finding arises from the fact that for arbitrary product states of the subsystems $A$ and $B$ any full correlation value $E_{AB}$ is the product of the corresponding marginal values:
\begin{equation}\label{eq:expValProd}
|\psi_{AB} \rangle = |\psi_A \rangle \otimes |\psi_B \rangle \Rightarrow E_{AB} = E_{A}E_{B}.
\end{equation}
This relation between single expectation values implies that the correlation distribution of parties $AB$ is a so-called product distribution of measurement results obtained on $A$ and $B$.
Whenever this is not the case we can infer that
the state is entangled across the partition $AB$.
Here, we first apply this criterion to pure product states, and later generalize it for arbitrary mixed states.

Consider first the triseparable state in Fig.~2a.
The bipartite distribution $E_{34}$, i.e., the distribution of the multiplication of outcomes for qubits $3$ and $4$, shows a logarithmic decay, which indicates a pure product state over these two parties.
The bipartite distribution $E_{12}$ is uniform as it is characteristic for maximally entangled two-qubit states.
The single qubit marginals confirm this observation:
$E_3$ and $E_4$ are almost uniform (pure states), whereas $E_1$ and $E_2$ correspond to the maximally mixed state.
Ideally, the correlation function for the maximally mixed state is equal to zero and results in a delta peak around $0$.
Finite statistics causes a broadening of this theoretical distribution and leads to the observed Gaussian shape.
Several of the distributions are product distributions.
For example we can verify that the full distribution $E_{1234}$ is the product distribution of multiplied results obtained on qubits $12$ and on qubits $34$,
and that $E_{34}$ is the product distribution of the results on qubit $3$ and on qubit $4$.
This is compatible with the state being separable across these partitions.
On the other hand, clearly the distribution $E_{12}$ is not a product one for the outcomes on qubit $1$ and on qubit $2$, which indicates the presence of entanglement.

The distributions for the biseparable state (\ref{eq:bisep}) are shown in Fig.~2b.
As expected, the bipartite marginal $E_{12}$ is the same as for the triseparable state.
The same also holds for the respective single qubit marginals of $E_1$ and $E_2$.
In the bipartite distribution of $E_{34}$, however, one can nicely observe the signature of a pure state intermediate between a maximally entangled and a product state, as tuned by the parameter $\varphi$.
For $\varphi\approx0.2$, the bipartite distribution of $E_{34}$ is almost uniform until approximately $0.5$ and decays logarithmically for larger values.
Equally, the respective single qubit marginals also show an intermediate behavior between a uniform distribution (pure state) until approximately $0.8$ and vanishing (white noise) for values above.
Both the distributions $E_{12}$ and $E_{34}$ do not correspond to the product distributions from the constituent subsystems which implies entanglement across these partitions of the pure state.

The maximally entangled GHZ state (Fig.~2c) and the cluster state (Fig.~2d) are not distinguishable on the level of the four respective single qubit marginals. 
Also certain bipartite marginals are the same, e.g., when tracing out qubits $3$ and $4$.
However, while for the permutationally invariant GHZ state all marginal distributions for the same number of qubits must be the same, a significantly different distribution (corresponding to the maximally mixed state) can be obtained for the cluster state, when tracing out for example qubits $1$ and $4$, i.e., for $E_{23}$.
Finally, the cluster and GHZ state can be distinguished also via their distributions of the full correlations.
From the plotted distributions for these two states only the distribution $E_{23}$ of the cluster state is (trivially) the product distribution for the results on qubits $2$ and $3$ (the same holds also for $E_{13}$, $E_{14}$, and $E_{24}$).
All other distributions are not the product distributions and thus reveal entanglement.

While our data reflect the theoretical predictions based on Eqs.~(\ref{EQ_STATES}a-d) well, there are systematic differences which can be traced back chiefly to a broadening of the distributions due to finite statistics~\cite{Knips2015}.
We used approximately $475$ counts per estimated expectation value for the GHZ state, giving rise to the broadening of a normal distribution with standard deviation on the order of $1/\sqrt{475}\approx0.046$. 
Accounting for these systematics is vital for the application of our quantitative analysis below and is explained in Methods.

\subsection*{Witnessing Entanglement}

To quantitatively analyze the experimentally obtained distributions, we focus on their statistical moments.
The $k$-th moment of the distribution of the full correlation is defined as
\begin{equation}
m_{1 \ldots n}^{(k)} \equiv \int_{SU(2)^n} {\rm d}U_1 \ldots {\rm d}U_n \; \tr{U^\dagger \sigma^{\otimes n}_z U \varrho}^k, 
\end{equation}
with $U\equiv U_1 \otimes \ldots \otimes U_n$ and where integration over $SU(2)$ is equivalent to sampling measurement directions uniformly from the single qubit Bloch spheres.
We will show in the following how to deduce the amount of purity and the presence of genuine multipartite entanglement using only the second moments of our measured correlation distributions.
We denote the second moment simply by $m_{1 \ldots n}\equiv m_{1 \ldots n}^{(2)}$.

One of the most elementary properties of a quantum state is its purity.
For $n$ qubits it reads (see Methods):
\begin{equation}\label{eq:purityN}
\Pur{\varrho}\equiv\tr{\varrho^2}=\frac{1}{2^n} \sum_{\setA \in \mathbb{P}(\setS)}{ 3^{|\setA|} \, m_\setA}
\end{equation}
where $\mathbb{P}(\setS)$ is the set of all subsets of $\setS=\{1,\ldots,n\}$ and $|\setA|$ denotes the cardinality of the set $\setA$.
Clearly, purity is accessible in the experiment with random measurements and forms the basis of our methods for detecting multipartite entanglement.
Note that in the case of a single qubit, the purity parameterizes the spectrum of the density matrix and hence any function of the quantum state which is invariant under local unitary transformations.

Let us consider the simplest case of pure two-qubit states.
The second moments of any product state satisfy $m_{12} = m_1 m_2$.
In consequence, the observation of $m_{12}>m_1 m_2$ indicates entanglement for pure states.
This reasoning cannot be easily extended to general states, since this inequality can also be satisfied for incoherent mixtures of product states.
However, we have found a purity dependent tightening of the inequality such that any $m_{12}$ above a certain purity dependent threshold must be due to quantum entanglement.
In Methods we derive the following entanglement witness condition:
\begin{equation}
{\mathcal M}_{2} \equiv m_{12} - m_1 m_2 \le 
\begin{cases}
4(1-\mathcal{P})\mathcal{P} / 9 & \text{\;for\;} \mathcal{P}\geq\frac{1}{2}, \\
(4\mathcal{P}-1)/9 & \text{\;for\;} \mathcal{P}<\frac{1}{2}.
\end{cases}
\label{EQ_2E}
\end{equation}
It holds for all separable states of two qubits with purity $\mathcal{P}\equiv\Pur{\varrho}$.
The bound is tight and achieved, e.g., by the state $p |00 \rangle \langle 00| + (1-p) |11 \rangle \langle 11 |$.
This powerful criterion can be generalized to the detection of genuine multipartite entanglement.

From the definition of genuine multipartite entanglement, i.e. entanglement which does not allow to represent a state as a mixture of product states across any bipartition, the left-hand side of Eq.~(\ref{EQ_2E}) generalizes for an $n$-qubit state to
\begin{equation}
{\mathcal M}_{n} \equiv m_{\setS} - \frac{1}{2} \sum_{\setA \in \{\mathbb{P}(\setS) \setminus (\setS \cup \varnothing)\}}{ m_\setA m_{\setS \setminus \setA}},
\label{EQ_nE}
\end{equation}
where the factor of $1/2$ resolves the issue of the double counting in the sum.

By numerical simulations, we find that the following condition holds for three-qubit bi-separable (not genuinely multipartite entangled) states
\begin{equation}
{\mathcal M}_{3} = m_{123} - m_1 m_{23} - m_2 m_{13} - m_3 m_{12} \le \tfrac{8}{27}(1-\mathcal{P})\mathcal{P}.
\label{EQ_3GEM}
\end{equation}
We have verified this inequality by extensive numerical search described in Methods.
The bound is tight for $\mathcal{P} \ge \frac{1}{2}$ and is achieved by, e.g., the state $p | \phi^+ \rangle \langle \phi^+| \otimes |0 \rangle \langle 0 | + (1-p) | \phi^- \rangle \langle \phi^-| \otimes |1 \rangle \langle 1 |$ 
with the Bell states $| \phi^{\pm} \rangle = \frac{1}{\sqrt{2}} (\ket{00} \pm \ket{11})$.

The bounds of the last two inequalities give hope for a simple dependence on the number of qubits.
A straightforward generalization from the previous bounds gives $(2/3)^4 (1-\mathcal{P})\mathcal{P}$. 
However, there exist bi-separable four-qubit states that violate this hypothetical bound.
We found by a numerical study that the inequality satisfied by bi-separable four qubit states has the following dependence on the purity,
\begin{eqnarray}
{\mathcal M}_{4}
\le \tfrac{8}{81}(1-\mathcal{P}^2).  \label{EQ_4GEM}
\end{eqnarray}
This bound is also tight for $\mathcal{P}\geq \frac{5}{8}$ and achieved by, e.g., the state $p | \phi^+ \rangle_{12} \langle \phi^+| \otimes | \phi^+ \rangle_{34} \langle \phi^+| + (1-p) | \phi^+ \rangle_{13} \langle \phi^+| \otimes | \phi^+ \rangle_{24} \langle \phi^+|$.

Our numerical results strongly indicate that any violation of inequality (\ref{EQ_3GEM}) or (\ref{EQ_4GEM}) certifies genuine multipartite entanglement between three or four qubits respectively.
We emphasize that these criteria require only the second moments of the observed distributions.

\begin{figure}
	\includegraphics[width=0.48\textwidth]{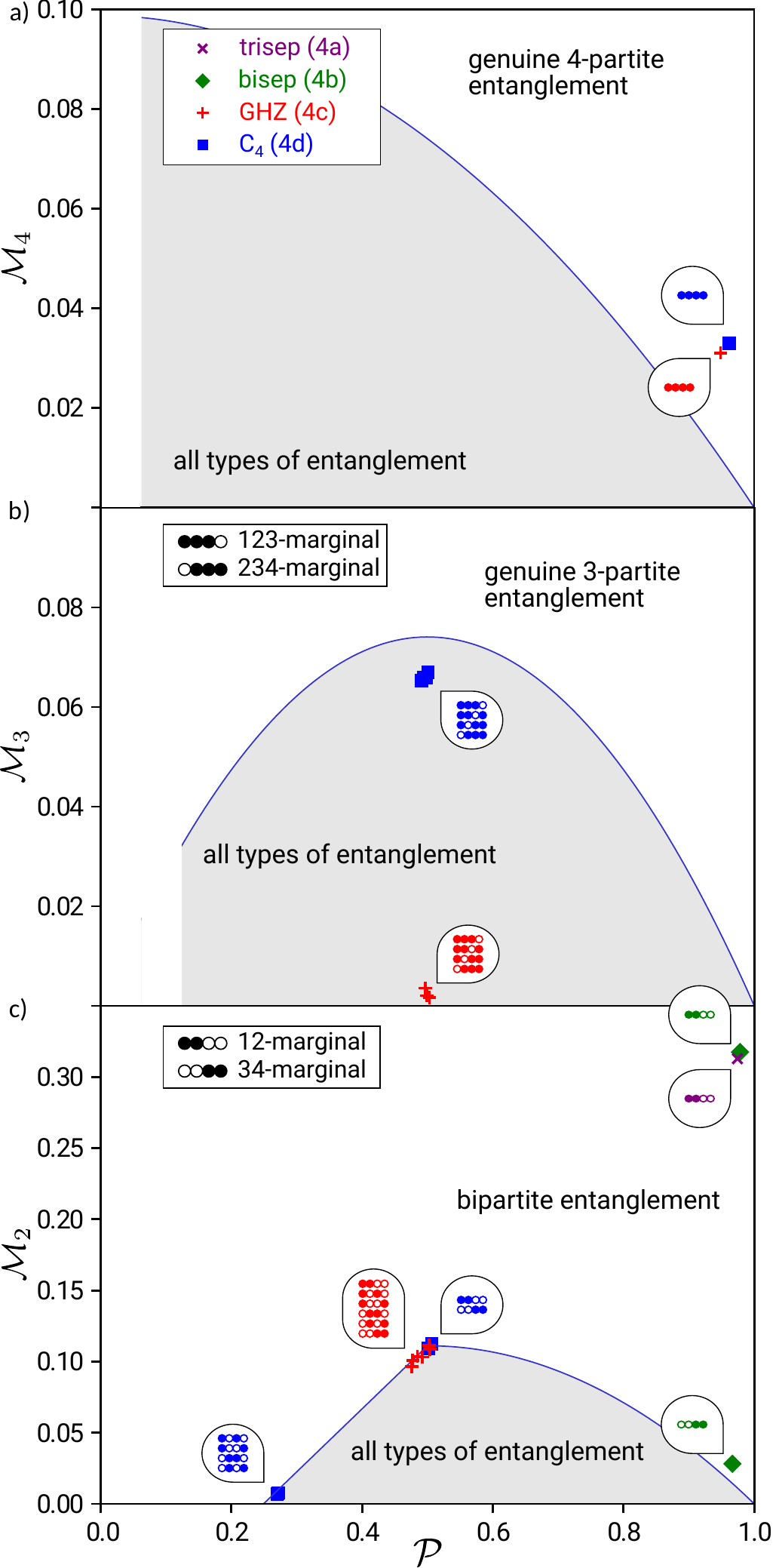}
	\caption{Analyzing the entanglement structure using $\mathcal{M}_n$: 
		(a) $\mathcal{M}_4$ of the GHZ state (\ref{eq:GHZ}) (red plus) and the cluster state (\ref{eq:cluster}) (blue square) are violating the bound for biseparable states (\ref{EQ_4GEM}), clearly indicating genuine $4$-partite entanglement.
		The negative values for $\mathcal{M}_4$ of the tri- and the biseparable states are not shown.
		(b) Evaluation of $\mathcal{M}_3$ for tripartite marginals for these states does not indicate any genuine tripartite entanglement as expected, as no point is found above the threshold given in Eq.~(\ref{EQ_3GEM}).
		The filled and non-filled circles indicate the type of marginals giving rise to different values of $\mathcal{M}_3$.
		(c) $\mathcal{M}_2$ is shown for all bipartite marginals.
		The four-qubit biseparable state (\ref{eq:bisep}) (green diamond) and the four-qubit triseparable state (\ref{eq:trisep}) (purple cross) have two and one marginals, respectively, which themselves are shown to be two-qubit entangled.
		The shaded regions contain all types of quantum states, irrespective of their entanglement properties.
		All error bars (standard deviations) are smaller than the markers.}
\end{figure}

Application of the conditions of Eqs.~(\ref{EQ_2E}), (\ref{EQ_3GEM}) and (\ref{EQ_4GEM}) to experimental data (Fig.~3) indeed enables detection of genuine $n$-partite entanglement for various subsets of particles.
For the cluster and the GHZ state, genuine $4$-partite entanglement is revealed with Eq.~(\ref{EQ_4GEM}) using $\mathcal{M}_4\approx0.0330\pm0.0004>0.0074\pm0.0012$ and $\mathcal{M}_4\approx0.0311\pm0.0006>0.0099\pm0.0012$, respectively.
The bi- and triseparable states do not violate their respective bound.
Investigating the entanglement properties for their marginal states, one can now prove the entanglement for the $12$-marginal and the $34$-marginal of the biseparable state as well as the $12$-marginal of the triseparable state.
It is therefore possible to conclude that the biseparable state contains contributions of at least $\varrho_{12}\otimes\varrho_{34}$, with entanglement between $1$ and $2$ and between $3$ and $4$, and the triseparable state contains $\varrho_{12}\otimes\varrho_{3}\otimes\varrho_{4}$. 
Note that the state could also contain genuine $4$-partite entanglement, which was not revealed by $\mathcal{M}_4$.

\section*{Discussion}

This work introduces a scheme to detect genuine multipartite entanglement and reveal its detailed structure in the absence of any reference frames and even for strongly fluctuating channels.
Key to this method is to subject a multipartite quantum system to randomly chosen local measurements and to analyze full and marginal correlations between local results using second moments of respective correlation distributions.
Haar random sampling removes any bias of the noise and, provided that the generation rate of multiqubit states is higher than the rate of fluctuations along the channel, neither the strength nor any characteristics of the noise matter.
The power of our procedure is demonstrated here by reconstructing the entanglement structure of various experimentally prepared photonic four-qubit states.
From this, many more interesting questions arise, e.g., whether it is possible to - up to suitable transformations - tomographically reconstruct quantum states or characterize quantum processes in our scenario of fully randomized local measurement directions.

\section*{Methods}

\subsection*{Finite sample size correction}
\label{sec:finitesample}

In our experiment, two different types of statistical effects have to be taken into account.
On one hand, for obtaining the distributions as in Fig.~2, a finite number $N_s$ of measurement settings ($N_s=10\,000$ in our case) is used.
This leads to an uncertainty in estimating the second moments $m_{\mathcal{A}}\equiv m_{\mathcal{A}}^{(2)}$.
This statistical error can be approximated by
\begin{equation}
\left(\Delta m_{\mathcal{A}}^{(2)}\right)^2 = \frac{1}{N_s}\left[m_{\mathcal{A}}^{(4)} - \frac{N_s-3}{N_s-1} \left(m_{\mathcal{A}}^{(2)}\right)^2\right],
\end{equation}
which describes the variance of the sample variance.

On the other hand, each correlation $E_{\mathcal{A}}^{(j)} \equiv E$ is obtained by performing $N_c$ measurements in the same setting.
Due to this finite sample size, for each expectation value in general we do not obtain the ideal result $E_R$, but measure a value $E_M$ at random from a conditional probability distribution $p(E_M|E_R)$,
approximately given by the Gaussian
\begin{equation}\label{binomGaussian}
p\left(E_M | E_R \right) = \frac{1}{\sqrt{2\pi}\sigma} \exp\left(-\frac{\left(E_M - E_R\right)^2}{2\sigma^2}\right)
\end{equation}
centered around $E_R$ with $\sigma=\sqrt{1-E_R^2}/\sqrt{N_c}$, see, e.g., \cite{Knips2015}.

This statistical deviation leads to an overestimation of $m_{\mathcal{A}}$.
We mitigate this systematic inaccuracy by taking into account the well known statistical effect from Eq.~(\ref{binomGaussian}).
Employing Bayesian methods, we are able to obtain $p\left(E_R | E_M \right)$ from $p\left(E_M | E_R \right)$ allowing to calculate $m_{\mathcal{A}}$ with reduced bias as
\begin{eqnarray}
m_{\mathcal{A}} &=& \int_{-1}^{1} {\rm d}E_R \, p(E_R) \, E^2_R \nonumber \\
&=& \int_{-1}^{1} {\rm d}E_R \; \int_{-1}^{1} {\rm d}E_M \, p(E_R|E_M) \, p(E_M) \; E^2_R.
\end{eqnarray}
Bayes' theorem provides $p(E_R|E_M)$ as
\begin{eqnarray}
p(E_R|E_M) & =& \frac{p(E_M|E_R) \tilde{p}(E_R)}{p(E_M)} \nonumber \\
& = &\frac{p(E_M|E_R) \tilde{p}(E_R)}{\int_{-1}^{1} {\rm d}E^\prime_R \, p(E_M|E^\prime_R) \, \tilde{p}(E^\prime_R)},
\end{eqnarray}
where $\tilde{p}(E_R)$ represents the prior assumption about the unknown distribution $p(E_R)$.
For our evaluation we use the measured distribution $p(E_M)$ as the prior guess about $p(E_R)$ and obtain an updated distribution according to the statistical analysis above.
This distribution is used to evaluate the moments.

\subsection*{Purity}

Per definition, the purity is $\mathcal{P}\equiv\tr{\varrho^2}$.
Any $n$-qubit state can be written as
\begin{equation}
\varrho = \frac{1}{2^n} \sum_{\mu_1 \dots \mu_n = 0}^3 T_{\mu_1 \dots \mu_n} \sigma_{\mu_1} \otimes \dots \otimes \sigma_{\mu_n},
\end{equation}
where $T_{\mu_1 \dots \mu_n} = \tr{\varrho \sigma_{\mu_1} \otimes \dots \otimes \sigma_{\mu_n}}$ and $\sigma$'s are the Pauli operators. Accordingly,
\begin{eqnarray}
\mathcal{P}&\equiv&\tr{\varrho^2}=\frac{1}{2^n}\sum_{\mu_1 \dots \mu_n = 0}^3 T_{\mu_1\mu_2\dots \mu_n}^2 \nonumber \\
&=&\frac{1}{2^{n}}\Bigg[T_{0\dots 0}^2 + \sum_{j_1 = 1}^3 T_{j_1 0 \dots 0}^2  + \dots + \sum_{j_n = 1}^3 T_{0 \dots 0 j_n}^2  \nonumber \\
& +& \sum_{j_1,j_2 = 1}^3 T_{j_1 j_2 0 \dots 0}^2 + \dots +\sum_{j_{n-1},j_n = 1}^3 T_{0 \dots 0 j_{n-1} j_n}^2 \nonumber \\
& +& \dots + \nonumber \\
& +& \sum_{j_1,\dots,j_n = 1}^3 T_{j_1\dots j_n}^2\Bigg] \nonumber\\
&=&\frac{1}{2^{n}}\Bigg[1 + 3 \left(m_1 + m_2 + \dots \right) +
3^2 \left(m_{12}+m_{13}+\dots \right) \nonumber \\
&& \;\; + \dots + 3^n m_{12\dots n}\Bigg] = \frac{1}{2^n} \sum_{\setA \in \mathbb{P}(\setS)}{ 3^{|\setA|} \, m_\setA},
\end{eqnarray}
where $\mathbb{P}(\setS)$ is the set of all subsets of $\setS=\{1,\ldots,n\}$ and $|\setA|$ denotes the cardinality of the set $\setA$, as in the main text.

\subsection*{Two-qubit condition}
\label{sec:proof2qubits}

Here we prove Eq.~(\ref{EQ_2E}) of the main text.
The problem is to maximize the value of $\mathcal{M}_2 = m_{12} - m_1 m_2$ over separable states of two qubits with a fixed purity $\mathcal{P}$.
Any two-qubit state admits a decomposition
\begin{equation}
\varrho = \frac{1}{4} \sum_{\mu,\nu = 0}^3 T_{\mu \nu} \sigma_{\mu} \otimes \sigma_{\nu},
\end{equation}
where $T_{\mu \nu} = \tr{\varrho \sigma_{\mu} \otimes \sigma_{\nu}}$.
In order to simplify numerical factors, we note that the second moments satisfy~\cite{PhysRevA.92.050301,PhysRevA.94.042302}:
\begin{eqnarray}
m_{12} & = & \frac{1}{9} \sum_{j,k=1}^3 T_{jk}^2 \equiv \frac{1}{9} \overline m_{12}, \\
m_{1} & = & \frac{1}{3} \sum_{j=1}^3 T_{j0}^2 \equiv \frac{1}{3} \overline m_{1}, \\
m_{2} & = & \frac{1}{3} \sum_{k=1}^3 T_{0k}^2 \equiv \frac{1}{3} \overline m_{2}.
\end{eqnarray}
The problem is therefore to maximize $\overline m_{12} - \overline m_1 \overline m_2$ (and then multiply the result by $\frac{1}{9}$).
Using the definition of the purity results in
\begin{equation}
\overline m_{12} = 4 \mathcal{P} - 1 - \overline m_1 - \overline m_2. 
\end{equation}
Therefore, the figure of merit reads:
\begin{eqnarray}
9 \mathcal{M}_2 & = & 4 \mathcal{P} - 1 - \overline m_1 - \overline m_2 - \overline m_1 \overline m_2  \nonumber \\
& \le & 4 \mathcal{P} - 1,
\end{eqnarray}
owing to the non-negativity of each second moment.
This bound holds for all states and is achieved by separable states of purity $\mathcal{P} \in [\frac{1}{4},\frac{1}{2}]$. 
An example for a state on the boundery is the mixture of white noise $\frac{1}{4} \openone$ with the classically correlated state $\frac{1}{2} | 00 \rangle \langle 00 | + \frac{1}{2} | 11 \rangle \langle 11 |$.

In order to derive the boundary for separable states with purity $\mathcal{P} \in [\frac{1}{2},1]$, we recall the definition of separability, i.e.
\begin{equation}
\varrho_{\mathrm{sep}} = \sum_j p_j \varrho_A^j \otimes \varrho_B^j.
\end{equation}
Therefore, any set of positive maps, but not necessarily completely positive, acting on a subsystem preserves separability.
Let us apply a so-called universal-not gate on subsystem $A$.
It is perhaps the simplest to introduce it using the Bloch sphere picture.
Universal-not reflects the Bloch vector of the state on which it acts about the origin, i.e. it is a linear map $\sigma_j \to - \sigma_j$ which puts a minus in front of every local Pauli operator. Clearly, any physical state, represented by the Bloch vector within a unit ball, is mapped to another physical state.
Yet, universal-not is not completely positive~\cite{Buzek1999}.
A generic two-qubit state is transformed by the universal-not gate on $A$ as follows:
\begin{eqnarray}
\varrho & = & \frac{1}{4} \Bigg( \openone \otimes \openone + \sum_{j = 1}^3 T_{j0} \sigma_j \otimes \openone + \sum_{k = 1}^3 T_{0k} \openone \otimes \sigma_k \nonumber \\
& + & \sum_{j,k=1}^3 T_{jk} \sigma_j \otimes \sigma_k \Bigg) \quad \to \nonumber \\
\overline \varrho & = & \frac{1}{4} \Bigg( \openone \otimes \openone - \sum_{j = 1}^3 T_{j0} \sigma_j \otimes \openone + \sum_{k = 1}^3 T_{0k} \openone \otimes \sigma_k \nonumber \\
& - & \sum_{j,k=1}^3 T_{jk} \sigma_j \otimes \sigma_k \Bigg).
\end{eqnarray}
Since we are assuming that $\varrho$ is separable, $\overline \varrho$ is also a separable physical state, i.e. a positive semi-definite operator.
Accordingly, the overlap between two positive semi-definite operators cannot be negative and we have
\begin{equation}
0 \le \tr{\varrho \overline \varrho} = \frac{1}{4}(1 - \overline m_1 + \overline m_2 - \overline m_{12}).
\end{equation}
Summing this up with the purity condition
\begin{equation}
\mathcal{P} = \frac{1}{4}(1 + \overline m_1 + \overline m_2 + \overline m_{12})
\label{EQ_2QP}
\end{equation}
gives the following inequality satisfied by all separable states with purity $\mathcal{P} $:
\begin{equation}
\overline m_2 \ge 2 \mathcal{P} - 1.
\label{SEP_2}
\end{equation}
By applying a universal-not on particle $B$ and following the same steps, one obtains
\begin{equation}
\overline m_1 \ge 2 \mathcal{P} - 1.
\label{SEP_1}
\end{equation}
Finally,
\begin{eqnarray}
\overline m_{12} - \overline m_1 \overline m_2 & = & 4 \mathcal{P} - 1 - \overline m_1 - \overline m_2 - \overline m_1 \overline m_2 \nonumber \\
& \le & 4 \mathcal{P} (1 - \mathcal{P}),
\label{EQ_NEW_CRIT}
\end{eqnarray}
where the inequality follows from (\ref{SEP_2}) and (\ref{SEP_1}).

\subsection*{Strength of the new criterion}

We now show that the new criterion, Eq.~(\ref{EQ_NEW_CRIT}), is stronger than those in Refs.~\cite{QuantInfComp.8.773,PhysRevA.77.062334,PhysRevA.80.042302,PhysRevA.80.042302,PhysRevA.90.042336,PhysRevA.92.050301,PhysRevA.94.042302},
which in the present notation read $\overline m_{12} \le 1$.
The underlying reason is that Eq.~(\ref{EQ_NEW_CRIT}) takes the purity and lower order correlations into account.

We first show that whenever $\overline m_{12} > 1$, then also our criterion is violated, i.e. 
$\overline m_{12} - \overline m_{1} \overline m_{2} - 4 \mathcal{P}(1 - \mathcal{P}) > 0$.
We start by rewriting the left-hand side of the latter using Eq.~(\ref{EQ_2QP}).
Next we utilise the condition $\overline m_{12} > 1$ in the resulting expression and this simplifies it to $\frac{1}{4}(\overline m_1 - \overline m_2)^2$, which is clearly non-negative.
In this context see also Ref.~\cite{HORO_BLOCH}, where an entanglement criterion is derived in terms of the difference between lengths of local Bloch vectors.

Finally, we present examples of entangled states for which $\overline m_{12} = 1$, but nevertheless the new criterion detects entanglement. 
For simplicity we represent the states in terms of the correlation tensor.
We choose $T_{xx} = T_{yy} = - T_{zz} = \frac{1}{\sqrt{3}}$, which ensures $\overline m_{12} = 1$, and also local Bloch vectors with $z$-components $T_{0z} = \frac{1}{6}(-3 + \sqrt{3} + \sqrt{2} \, 3^{3/4})$ and $T_{z0} = \frac{1}{6}(-3 + \sqrt{3} - \sqrt{2} \, 3^{3/4})$.
Among many physically allowed values of $T_{0z}$ and $T_{z0}$ for which the new criterion is violated, the ones given here produce maximal violation.

\subsection*{Numerical simulations}
\label{sec:NumSim}

\begin{figure}
	\includegraphics[width=0.48\textwidth]{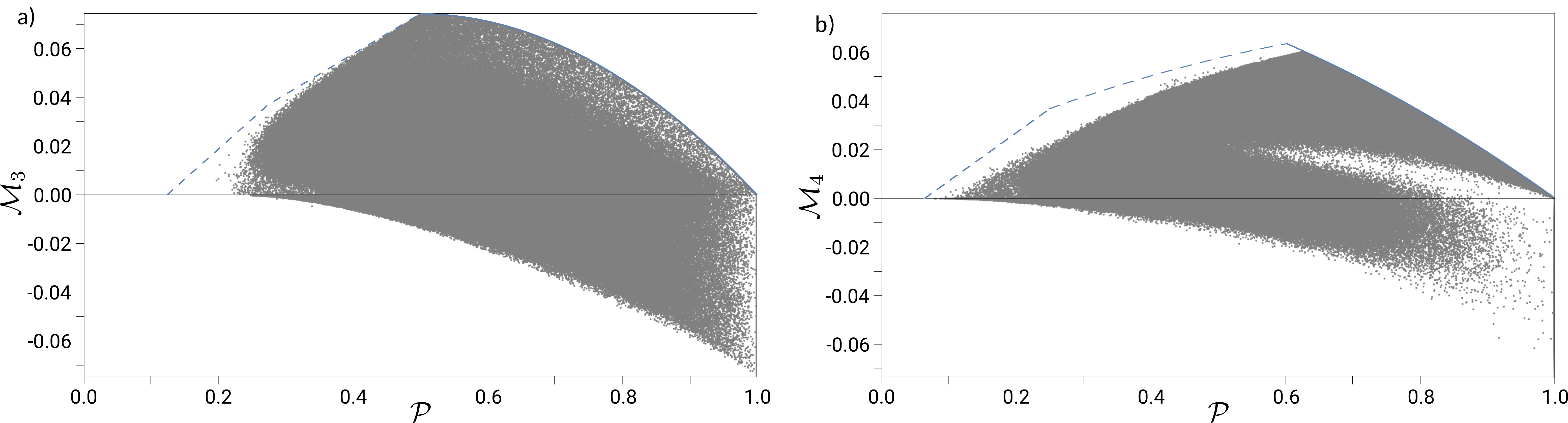}
	\caption{ Numerical evidence supports our witnesses of genuine tripartite and four-partite entanglement.
		We sampled more that $10^6$ biseparable states from various (also random) families.
		The numerical boundary for biseparable states is plotted with a solid line, whereas the numerical boundary that holds for all quantum states (boundary of physicality) is plotted as a dashed line.
		(a) The biseparable states of 3 qubits are confined to the region below the boundary given by Eq.~(\ref{EQ_M3BOUND}).
		(b) The biseparable states of 4 qubits are confined to the region below the boundary given by $\frac{8}{81}(1-\mathcal{P}^2)$. }
\end{figure}

Here we give numerical evidence for the bounds of Eqs.~(\ref{EQ_3GEM}) and (\ref{EQ_4GEM}) of the main text.
We performed sampling of more than $10^6$ biseparable states and always found the bounds satisfied.
Fig.~4 illustrates the results of numerical simulation. 

For the case of three qubits we find the following improved boundary for small values of $\mathcal{P}$:
\begin{equation}
\label{EQ_M3BOUND}
{\mathcal M}_{3} \le
\begin{cases}
(8\mathcal{P}-1) / 27 & \text{\;for\;} \mathcal{P} \in [\frac{1}{8}, \frac{1}{4}], \\
4 \mathcal{P} / 27 & \text{\;for\;} \mathcal{P} \in (\frac{1}{4}, \frac{1}{2}], \\
8 (1-\mathcal{P})\mathcal{P} / 27 & \text{\;for\;} \mathcal{P}>\frac{1}{2},
\end{cases}
\end{equation}
while the improved boundary for four qubits reads
\begin{equation}
\label{EQ_M4BOUND}
{\mathcal M}_{4} \le
\begin{cases}
(16\mathcal{P}-1) / 81 & \text{\;for\;} \mathcal{P} \in [\frac{1}{16}, \frac{1}{4}], \\
2 (-8\mathcal{P}^2+16\mathcal{P}+1) / 243 & \text{\;for\;} \mathcal{P} \in (\frac{1}{4}, \mathcal{P}_0], \\
8 (1-\mathcal{P}^2) / 81 & \text{\;for\;} \mathcal{P}>\mathcal{P}_0,
\end{cases}
\end{equation}
where $\mathcal{P}_0=\frac{-4+3\sqrt{3}}{2}\approx0.60$.

\subsection*{Acknowledgements}

We thank Otfried G\"uhne, Felix Huber and Nikolai Wyderka for fruitful discussions.
This research was supported by the DFG (Germany) and NCN (Poland) within the joint funding initiative ``Beethoven 2'' (2016/23/G/ST2/04273, 381445721), 
and by the DFG under Germany's Excellence Strategy EXC-2111 390814868.
We acknowledge the Singapore Ministry of Education Academic Research Fund Tier 2, Project No. MOE2015-T2-2-034.
WL acknowledges partial support from the Foundation for Polish Science (IRAP project ICTQT, Contract No. 2018/MAB/5, cofinanced by EU via Smart Growth Operational Programme).
TP is supported by the Polish National Agency for Academic Exchange NAWA Project No. PPN/PPO/2018/1/00007/U/00001.
JD and LK acknowledge support from the PhD programs IMPRS-QST and ExQM, respectively.



\end{document}